\documentstyle[12pt,draft]{article}
\newcounter{dummy}{}
\newcommand{\letters}{\setcounter{dummy}{\value{equation}}
\refstepcounter{dummy}
\setcounter{equation}{0}
\renewcommand{\theequation}{\arabic{dummy}\alph{equation}}}
\newcommand{\noletters}{\setcounter{equation}{\value{dummy}}
\renewcommand{\theequation}{\arabic{equation}}}
\newenvironment{mathletters}{\letters}{\noletters}

\begin{document}

\title{Multiinstantons in curvilinear coordinates.}
\author{A.~A.~Abrikosov,~Jr.\thanks{E-mail:
{\sc persik@vxitep.itep.ru}} \\
{\em 117 259, Moscow, ITEP.}}
\date{}
\maketitle

\begin{abstract}
The 'tHooft's $5N$-parametric multiinstanton solution is generalized to
curvilinear coordinates. Expressions can be simplified by the gauge
transformation that makes $\eta$-symbols constant in the vierbein
formalism.  This generates the compensating addition to the gauge
potential of pseudoparticles. Typical examples (4-spherical, $2+2$- and
$3+1$-cylindrical coordinates) are studied and explicit formulae presented
for reference.  Singularities of the compensating field are discussed.
They are irrelevant for physics but affect gauge dependent quantities.
\end{abstract}

\hyphenation{pseu-do-par-ticle pseu-do-par-ticles}

\section*{Introduction}

Instantons form an essential nonperturbative element of nonabelian gauge
theories applied in physics of strong and weak interactions. But despite
a significant knowledge that was acquired since the pioneering paper
\cite{BPST} the topic is not exhausted yet. Among others the question of
the role of instantons in quark confinement remains unsettled.
Probably the problem will not be solved unless new methods and ideas
appear. Reviews of the current state of affairs can be found in
\cite{Shifman,Schaefer/Shuryak}.

Usually instantons are studied in the infinite 4-dimensional euclidean
space pa\-ra\-met\-rized by Cartesian coordinates. A demand for
non-Cartesian coordinates first came from investigation of anomalies and
index theorems on manifolds with boundaries. Adaptation of existing
results (see \cite{Wipf,NiSe}) to practical problems requires explicit
formulae.

A promising direction would be to examine pseudoparticles in confinement
models such as the QCD-string or a MIT-bag. However Cartesian coordinates
are not the best choice for such objects. For example strings looks more
natural in $(2+2)$-cylindrical coordinates while a $(3+1)$-frame may be
convenient for the bag. Studies of pseudoparticles against nonuniform
backgrounds may require even more imagination.

A notable progress in the instanton physics was associated with exact
multiinstanton solutions. The general solution \cite{AHDM} proved too
complicated and many physicists preferred the convenient 'tHooft's {\em
Ansatz\/} \cite{5N}. Most of instanton-based models start from
pseudoparticles in singular gauge that are the specific case of this
solution. An advantage of the {\em Ansatz\/} is that it makes possible to
calculate exact Green functions in multipseudoparticle field \cite{BCCL}.
Besides it provides a way to study QCD at nonzero temperature
\cite{HS,GYP} and quark density \cite{Abrikosov}.

The purpose of the present paper is to develop a simple approach to
'tHooft's multipseudoparticle solution in curvilinear coordinates. This
offers a chance to benefit from it's advantages preserving spatial
symmetry of other objects. This would simplify calculations and widen the
research field.

A basic component of the Cartesian 'tHooft's solution are the so-called
$\eta $-symbols ($\eta $-tensors) that project {\em (anti-)\/} selfdual
tensors onto the $SU(2)$ gauge group, \cite{tHooft}. Being transformed
to other coordinates $\eta $-symbols loose their simple form.  Our idea is
to consider instead of them the $\xi $-symbols that are literal analogues
of $\eta $'s in the vierbein formalism. The $\xi $-symbols can be made
constant by a gauge transform related to the metric tensor. The resulting
$5N$-parametric gauge potential is the sum of the 'tHooft's multiinstanton
and the compensating field. The latter may be found by a simple
calculation.  After discussing general aspects we illustrate the idea by
an example and derive explicit formulae for two typical cases. It turns
out that the compensating potential may exhibit singularities that affect
gauge dependent quantities.

The paper has the following structure. We start from reminding the basics:
Sect.~\ref{instantons} introduces the 'tHooft's {\em Ansatz\/} while
Sect.~\ref{coordinates} deals with curvilinear coordinates. Section~\ref
{comp_field} is dedicated to the compensating gauge connection. First we
calculate the compensating gauge potential and then discuss the relation
between $\xi $- and $\eta $-symbols. After that we turn to pseudoparticles
in non-Cartesian frame in Sect.~\ref{inst+comp}. We calculate vector
potentials and gauge field strengths for a general $N$-instanton
configuration and for an instanton in singular and regular gauges%
\footnote{ More precisely our solutions coincide with those in Cartesian
coordinates}.

Section~\ref{Sect_O(4)} presents a detailed study of an instanton in
$4$-spherical coordinates. We calculate explicitly the (gauged)
pseudoparticle fields in singular and regular gauges and obtain the same
result provided that the instanton is at the origin. This may serve an
indication in favour of the approach. The calculation of the Chern-Simons
number reveals that being gauge dependent it feels the singularity of the
compensating field.  In Section~\ref{cylinder} we compute the compensating
connection for $(2+2)$- and $(3+1)$-cylindrical coordinates. The results
are summarized in the Conclusion.

\section{Instantons in singular and regular gauges.}


\label{instantons}This section reviews basic facts on instantons in
4-dimensional euclidean space. For time being we shall focus on Cartesian
coordinates and make no distinction between upper and lower indices. We
shall consider a Yang-Mills theory with the $SU(2)$ gauge group. The action
of the gauge field is:
\begin{equation}
S =\int d^4x\,\frac{\,\,(F_{\mu \nu }^a)^2}{4g^2}=\frac 1{4g^2}\int
d^4x\,(\partial _\mu A_\nu ^a-\partial _\nu A_\mu ^a+\epsilon ^{abc}A_\mu
^bA_\nu ^c)^2.  \label{s_gauge}
\end{equation}
Vector potential is denoted by $A_\mu ^a$ with $\mu =1,\ldots ,4$ and
$a=1,2,3$ being the Lorentz and group indices respectively. Throughout the
paper we shall use for gauge fields the matrix notation. The covariant
derivative in fundamental representation is:
\begin{equation}
D_\mu =\partial _\mu - i\hat{A}_\mu
=\partial _\mu -\frac i2\tau ^aA_\mu ^a,
\end{equation}
where $\tau ^a$ are Pauli matrices and hats indicate matrices. Later we
shall introduce the Levi-Civita and spin connections that are absent in
Cartesian coordinates. Commutator of the covariant derivatives gives the
field strength:
\begin{equation}
\hat{F}_{\mu \nu }=\frac 12\tau ^a\,F_{\mu \nu }^a=i\,\left[ D_\mu ,\,D_\nu
\right] .  \label{[DmuDnu]}
\end{equation}

Classical equations for $A_\mu ^a$ have instanton, or pseudoparticle,
solutions. Instantons are characterized by their topological properties:
their field is selfdual, (a), and has the unit topological charge, (b):
\begin{equation}
\hat{F}_{\mu \nu }^I=\frac 12\epsilon _{\mu \nu \lambda \sigma }
\hat{F}_{\lambda \sigma }^I,\quad {\rm (a)};\qquad \frac 1{32\pi ^2}\int
d^4x\,\epsilon _{\mu \nu \lambda \sigma }\,\mbox{\rm tr}\,\hat{F}_{\mu \nu
}^I\,\hat{F}_{\lambda \sigma }^I=1,\quad {\rm (b)}.  \label{tp_chrg}
\end{equation}

In addition to instantons there exist anti-instantons. Anti-instanton field
$\hat{F}_{\mu \nu }^A$ is antiselfdual and has the topological charge $-1$,
{\em i. e.} the signs of the right hand side of the equations (\ref{tp_chrg})
must be reversed. From here on we shall speak mostly about instantons
indicating generalization to anti-instantons if necessary.

Instanton field depends on gauge. The two most popular are regular and
singular gauges. The fields of instantons of radius $\rho $ centered at
$x^0$ are (up to uniform gauge rotations):
\begin{mathletters}
\label{r/s_matr}
\begin{eqnarray}
\left. \hat A_\mu ^{+}\right| _{{\rm reg}} &=
&\frac{\hat{\eta}_{\mu \nu }^{+}}2\partial _\nu
\ln \Pi _{{\rm reg}}(x,\,x^0)=
\frac{\hat{\eta}_{\mu \nu }^{+}}2\partial _\nu \ln
\left[ \left( x-x^0\right) ^2+\rho ^2\right] ;
\label{r/s_matra} \\
\left. \hat A_\mu ^{+}\right| _{{\rm sing}} &=
&-\frac{\hat{\eta}_{\mu \nu }^{-}}2\partial _\nu
\ln \Pi _{{\rm sing}}(x,\,x^0)=-\frac{\hat{\eta}_{\mu \nu
}^{-}}2\partial _\nu \ln \left[ 1+\frac{\rho ^2}{\left( x-x^0\right) ^2}
\right] .  \label{r/s_matrb}
\end{eqnarray}
Here $\hat{\eta}_{\mu \nu }^{+}=\tau ^a\eta _{\mu \nu }^a$ and
$\hat{\eta}_{\mu \nu }^{-}=\tau ^a\bar{\eta}_{\mu \nu }^a$ are the matrix
versions of the 'tHooft's $\eta $-symbols defined by:
\end{mathletters}
\begin{equation}
\eta (\bar{\eta})_{\mu \nu }^a=-\eta (\bar{\eta})_{\nu \mu }^a=\left\{
\begin{array}{ccc}
\epsilon ^{a\mu \nu } & \quad {\rm for}\quad & \mu ,\nu =1,2,3; \\
(-)\delta ^{\mu a} & \quad {\rm for}\quad & \nu =4.
\end{array}
\right.
\end{equation}
The $\eta $-symbols are antisymmetric in Lorentz indices and selfdual or
antiselfdual respectively, {\em i.~e.\/}
\begin{equation}
\frac 12\epsilon _{\mu \nu \lambda \sigma }\hat{\eta}_{\lambda \sigma }^{+}=
\hat{\eta}_{\mu \nu }^{+};\qquad \frac 12\epsilon _{\mu \nu \lambda \sigma }
\hat{\eta}_{\lambda \sigma }^{-}=-\hat{\eta}_{\mu \nu }^{-}.
\end{equation}
Their properties can be found in \cite{tHooft} or obtained directly from
the matrix representation (\ref{tau_mu}) below.

In practical calculations it is convenient to make use of the two hermitean
conjugated sets of $2\times 2$ matrices (latin indices stand for the three
``spatial'' dimensions):
\begin{equation}
\tau _\mu =(\tau ^a,i);\qquad \tau _\mu ^{\dagger }=(\tau ^a,-i).
\label{tau,tau^dag}
\end{equation}
The following equations relate $\tau _\mu ^{\vphantom{\dagger}}$, $\tau _\mu
^{\dagger }$ to $\hat \eta $-symbols:
\begin{equation}
\tau _\mu ^{\dagger }\tau _\nu ^{\vphantom{\dagger}}=\delta _{\mu \nu }+i
\hat{\eta}_{\mu \nu }^{+};\qquad \qquad \tau _\mu ^{\vphantom{\dagger}}\tau
_\nu ^{\dagger }=\delta _{\mu \nu }+i\hat{\eta}_{\mu \nu }^{-};
\label{tau_mu}
\end{equation}
Separating the symmetric and antisymmetric parts of these expressions we
obtain:
\begin{equation}
\tau _{\{\mu }^{\vphantom{\dagger}}\tau _{\nu \}}^{\dagger }=\tau _{\{\mu
}^{\dagger }\tau _{\nu \}}^{\vphantom{\dagger}}=\delta _{\mu \nu };\qquad
\tau _{[\mu }^{\dagger }\tau _{\nu ]}^{\vphantom{\dagger}}=i\hat{\eta}_{\mu
\nu }^{+};\qquad \tau _{[\mu }^{\vphantom{\dagger}}\tau _{\nu ]}^{\dagger }=
i \hat{\eta}_{\mu \nu }^{-}.  \label{t-comm}
\end{equation}

Expression (\ref{r/s_matrb}) can be obtained from (\ref{r/s_matra}) by means
of the gauge transform:
\begin{equation}
\left. \hat{A}_\mu ^I(x)\right| _{{\rm sing}}=\hat{N}_{+}^{-1}\left.
\hat{A}_\mu ^I(x)
\right| _{{\rm reg}}\hat{N}_{+}+i\,\hat{N}_{+}^{-1}\,\partial _\mu
\hat{N}_{+},  \label{A_s=OA_rO+OdO}
\end{equation}
with the matrices $\hat{N}_{+}=x_\lambda \tau _\lambda ^{\dagger }/x$ and
$\hat{N}_{+}^{-1}=x_\lambda \tau _\lambda ^{\vphantom{\dagger}}/x$ ($\tau $
and $\tau ^{\dagger }$ being swapped for anti-instantons,
$\hat{N}_{-}=\hat{N}_{+}^{-1}$).

Another way to derive (\ref{r/s_matrb}) is to carry out the inversion,
$x_\mu \rightarrow x_\mu \,\rho ^2/x^2$. However the latter changes the
orientation of the coordinate system and converts instanton to
anti-instanton. This is corrected by the $\hat{\eta}^{\pm }\rightarrow
\hat{\eta}^{\mp }$ replacement.

Singular gauge has several practical advantages. Gauge potentials fall
rapidly and pseudoparticles are almost independent. Generalization to
multipseudoparticle configurations is done simply by adding extra pieces to
$\Pi (x)$:
\begin{equation}
\Pi _{{\rm sing}}(x)\rightarrow \Pi _N(x)=
1+\sum_{i=1}^N\frac{\rho _i^2}{(x-x_i)^2},
\end{equation}
$\rho _i$ and $x_i$ being radii and positions of pseudoparticles. This
solution bears the name of the 'tHooft's {\em Ansatz,\/} \cite{5N}.
It depends on $5N$ parameters $\{x_i,\,\rho _i\}$ and does not
allow for independent gauge rotations of instantons. But the lack of
generality is balanced by the extreme handiness.

\section{Curvilinear coordinates.}

\label{coordinates}In order to distinguish curvilinear coordinates from the
Cartesian ones we shall denote them by $q^\alpha $, $q^\beta $ {\em etc.\/}
The metric tensor is now $g_{\alpha \beta }(q)$,
\begin{equation}
ds^2 = dx^2_\mu = g_{\alpha \beta }(q)\,dq^\alpha \,dq^\beta .
\end{equation}
The metric $g_{\alpha \beta }(q)$ and it's inverse $g^{\alpha \beta
}(q)=\left[ g_{\alpha \beta }(q)\right] ^{-1}$ are used for raising and
lowering indices: $A_\alpha =g_{\alpha \beta }A^\beta $; $A^\alpha
=g^{\alpha \beta }A_\beta $.

Let us start from fields that are singlet with respect to the gauge group.
Covariant derivatives of vectors are taken with the help of the Levi-Civita
connection $\Gamma _{\beta \gamma }^\alpha $:
\begin{equation}
D_\alpha A^\beta =\partial _\alpha A^\beta +\Gamma _{\alpha \delta }^\beta
A^\delta ;\qquad {\rm and}\qquad D_\alpha A_\beta =\partial _\alpha A_\beta
-\Gamma _{\alpha \beta }^\delta A_\delta ;
\end{equation}
The latter is unambiguously defined by the condition that the metric tensor
is covariantly constant, $D_\alpha g_{\beta \gamma }=0$.
\begin{equation}
\Gamma _{\beta \gamma }^\alpha =\frac 12g^{\alpha \delta }\left(
\frac{\partial g_{\delta \beta }}{\partial q^\gamma }+
\frac{\partial g_{\delta
\gamma }}{\partial q^\beta }-\frac{\partial g_{\beta \gamma }}{\partial
q^\delta }\right) .  \label{Gamma}
\end{equation}

The metric $g_{\alpha \beta }$ may be decomposed into vierbeins $e_\alpha ^a$
(from here on we reserve latin indices for the latter). Let $\pi _{ab}$ be
the flat euclidean metric, $\pi ^{ab}=\pi _{ab}={\rm
diag\,}(1,\,1,\,1,\,1)$. Then
\begin{equation} g_{\alpha \beta }(q)=\pi
_{ab}\,e_\alpha ^a(q)\,e_\beta ^b(q);\qquad {\rm  while}\qquad \pi
^{ab}=g^{\alpha \beta }(q)\,e_\alpha ^a(q)\,e_\beta ^b(q).  \label{g<=>pi}
\end{equation}
One may convert spatial indices into vierbein ones, $A^a=e_\alpha ^aA^\alpha
$. Raising and lowering of the latter is performed by means of the tensors
$\pi $: $A_a=\pi _{ab}A^b$ and $A^a=\pi ^{ab}A_b$. The inverse of the
vierbein is $e_a^\alpha $:
\begin{equation}
e_\alpha ^a\,e_b^\alpha =\delta _b^a;\qquad {\rm and}\qquad e_a^\alpha
\,e_\beta ^a=\delta _\beta ^\alpha .
\end{equation}

Covariant derivatives of quantities with vierbein indices are defined in
terms of the spin connection $R_{\alpha \,b}^a$.
\begin{equation}
D_\alpha A^a=\partial _\alpha A^a+R_{\alpha \,b}^a\,A^b;\qquad {\rm and}
\qquad D_\alpha A_a=\partial _\alpha A_a-A_b\,R_{\alpha \,a}^b.
\end{equation}
Vierbeins are covariantly constant and this fixes the spin connection.
Solving the equation $D_\alpha e_a^\beta =\partial _\alpha e_a^\beta +\Gamma
_{\alpha \gamma }^\beta \,e_a^\gamma -e_b^\beta \,R_{\alpha \,a}^b=0$ we
obtain:
\begin{equation}
R_{\alpha \,b}^a=e_\beta ^a\,\partial _\alpha e_b^\beta +e_\beta ^a\,\Gamma
_{\alpha \gamma }^\beta \,e_b^\gamma =e_\beta ^a\,(D_\alpha e^\beta )_b.
\label{R_alpha}
\end{equation}
The matrices $R$ are antisymmetric with respect to the exchange
$a\leftrightarrow b$.
(This follows from $e_\mu ^ae_b^\mu =\delta _b^a={\it
const.}$\/) Sometimes it is convenient to expand $R_{\alpha \,b}^a$ in terms
of generators of the $O(4)$ group, $(L_{mn})_b^a$:
\begin{equation}
R_{\alpha \,b}^a=-\frac i2B_\alpha ^{mn}\,(L_{mn})_b^a,\qquad {\rm where}
\qquad (L_{mn})_b^a=-i\left( \delta _m^a\,\pi _{bn}-\delta _n^a\,\pi
_{bm}\right) .  \label{B_alpha}
\end{equation}
The coefficients $B_\alpha ^{mn}$ are antisymmetric, $B_\alpha
^{mn}=-B_\alpha ^{nm}$,
\begin{equation}
B_\alpha ^{mn}=\frac i2{\rm tr\,}\hat{R}_\alpha \,\hat{L}^{mn}=\frac i2\,
(\hat{R}_\alpha )_b^a\,(\hat{L}^{mn})_a^b.
\end{equation}

The last thing is to extend the spin connection and covariant derivative
to spin-$\frac 12$ fields. We would like matrices $\gamma _a$ with latin
indices to be covariantly constant. It is easy to see that if we define
$\gamma $-matrices and their commutators so that
\begin{equation}
\left\{ \gamma _a,\,\gamma _b\right\} =2\delta _{ab},\qquad {\rm and}\qquad
\sigma _{ab}=-\frac i2\left[ \gamma _a,\,\gamma _b\right] ,
\label{[g_a,g_b]}
\end{equation}
then the matrices $\sigma _{mn}/2$ are rotation generators for spin-$\frac
12 $ fields and
\begin{equation}
D_\alpha \gamma _a=\partial _\alpha \gamma _a+\frac i2B_\alpha
^{mn}\,\left\{ \gamma _b(L_{mn})_a^b-\frac 12\left[ \sigma _{mn},\,\gamma
_a\right] \right\} =0.  \label{Dgamma_a}
\end{equation}
The last relation can be easily checked directly.

\section{The compensating gauge connection\label{comp_field}}

\subsection{Definition and properties of the compensating connection}

Now let us turn to gauge fields in curvilinear coordinates. One may define
$\tau _a^{\vphantom{\dagger}}$ and $\tau _a^{\dagger }$ matrices with
vierbein indices by analogy with (\ref{tau,tau^dag}). We shall show that it
is possible to introduce a compensating gauge field such that either $\tau
_a^{\vphantom{\dagger}}\tau _b^{\dagger }$ or $\tau _a^{\dagger }
\tau _b^{\vphantom{\dagger}}$ is covariantly constant. The compensating
field turns out to be a pure gauge provided that the space is flat.

It is convenient to turn back to the spin connection $B_\alpha ^{mn}$ and
use the following representation of $\gamma $-matrices, (check \ref
{[g_a,g_b]}):
\begin{equation}
\gamma _a=\left(
\begin{array}{cc}
0 & \tau _a^{\vphantom{\dagger}} \\
\tau _a^{\dagger } & 0
\end{array}
\right) ;\qquad \gamma _a\,\gamma _b=\left(
\begin{array}{cc}
\tau _a^{\vphantom{\dagger}}\tau _b^{\dagger } & 0 \\
0 & \tau _a^{\dagger }\tau _b^{\vphantom{\dagger}}
\end{array}
\right) ;\qquad \sigma _{ab}=\left(
\begin{array}{cc}
\hat{\xi}_{ab}^{-} & 0 \\
0 & \hat{\xi}_{ab}^{+}
\end{array}
\right) ;  \label{[g,g]}
\end{equation}
where
\begin{equation}
\hat{\xi}_{ab}^{+}=-i\,\tau _{[a}^{\dagger }
\tau _{b]}^{\vphantom{\dagger}
}\qquad {\rm and}\qquad \hat{\xi}_{ab}^{-}=-i\,
\tau _{[a}^{\vphantom{\dagger}
}\tau _{b]}^{\dagger }  \label{xi_ab}
\end{equation}
are the vierbein analogues of the Cartesian $\hat{\eta}_{\mu \nu }^{+}$
and $\hat{\eta}_{\mu \nu }^{-}$. One may define $\hat \xi$-symbols with
coordinate indices as follows:
\begin{equation}
\hat{\xi}_{\alpha \beta }^{+}=e_\alpha ^a\,e_\beta ^b\,
\hat{\xi}_{ab}^{+}\qquad {\rm and}\qquad \hat{\xi}_{\alpha \beta }^{-}
=e_\alpha^a\,e_\beta ^b\,\hat{\xi}_{ab}^{-}.  \label{xi_alphabet}
\end{equation}
In order to escape confusion we shall use only $\hat{\xi}_{ab}$ that are
just constant numerical matrices. From the relation (\ref{Dgamma_a}) applied
to the block-diagonal matrix $\gamma _a\,\gamma _b$ it is easy to deduce for
the separate blocks that:
\begin{mathletters}
\label{Dxi^pm}
\begin{eqnarray}
D_\alpha \,\tau _a^{\dagger }\tau _b^{\vphantom{\dagger}} &=&\partial
_\alpha \,\tau _a^{\dagger }\tau _b^{\vphantom{\dagger}}+\frac i2B_\alpha
^{mn}\,\tau _c^{\dagger }\tau _d^{\vphantom{\dagger}}\left( \delta
_a^c\,(L_{mn})_b^d+(L_{mn})_a^c\,\delta _b^d\right) -i\,\left[
\hat{A}_\alpha ^{+},\,
\tau _a^{\dagger }\tau _b^{\vphantom{\dagger}}\right] =0,
\label{Dxi^pma} \\
D_\alpha \,\tau _a^{\vphantom{\dagger}}\tau _b^{\dagger } &=&\partial
_\alpha \,\tau _a^{\vphantom{\dagger}}\tau _b^{\dagger }-\frac i2B_\alpha
^{mn}\,\tau _c^{\vphantom{\dagger}}\tau _d^{\dagger }\left( \delta
_a^c\,(L_{mn})_b^d+(L_{mn})_a^c\,\delta _b^d\right) -i\left[ \hat{A}_\alpha
^{-},\,\tau _a^{\vphantom{\dagger}}\tau _b^{\dagger }\right] =0,
\label{Dxi^pmb}
\end{eqnarray}
\end{mathletters}
where the vector-potentials $\hat{A}_\alpha ^{+}$ and $\hat{A}_\alpha ^{-}$
are:
\begin{mathletters}
\label{A^pm}
\begin{eqnarray}
\hat{A}_\alpha ^{+} &=&\frac 14B_\alpha ^{mn}\,\hat{\xi}_{mn}^{+}=\frac
i4\tau _a^{\dagger }\,R_\alpha ^{ab}\,\,\tau _b^{\vphantom{\dagger}};
\label{A^pma} \\
\hat{A}_\alpha ^{-} &=&\frac 14B_\alpha ^{mn}\,\hat{\xi}_{mn}^{-}=\frac
i4\tau _a^{\vphantom{\dagger}}\,R_\alpha ^{ab}\,\tau _b^{\dagger };
\label{A^pmb}
\end{eqnarray}
\end{mathletters}

Thus the gauge connections $\hat{A}_\alpha ^{+}$ and $\hat{A}_\alpha ^{-}$
are projections of the selfdual and antiselfdual parts of the spin
connection $R_\alpha ^{ab}$ onto the $SU(2)$ gauge group. In general only
one of the bilinears $\tau _a^{\dagger }\tau _b^{\vphantom{\dagger}}$, $\tau
_a^{\vphantom{\dagger}}\tau _b^{\dagger }$ can be made covariantly constant
by the appropriate choice of the compensating field. The equations (\ref
{A^pm}) are the most straightforward way to calculate $\hat{A}^{\pm }$.
(Note that together with $\hat{\xi}_{ab}^{\pm }$ the corresponding
$\hat{\xi}_{\alpha \beta }^{\pm }$, (\ref{xi_alphabet}) becomes
covariantly constant as well.)

In flat euclidean space either of the potentials $\hat{A}_\alpha ^{+}$ and
$\hat{A}_\alpha ^{-}$ is the pure gauge, {\em i.~e.\/} the corresponding
field strengths $\hat{F}_{\alpha \beta }^{\pm }=\partial _\alpha
\hat{A}_\beta ^{\pm }-\partial _\beta A_\alpha ^{\pm }-
i\left[ A_\alpha ^{\pm },\,
\hat{A}_\beta ^{\pm }\right] $ are zero. In order to show that let us again
resort to the help of the spin connection. Note that matrices $\gamma
_\alpha =e_\alpha ^a\,\gamma _a$ are covariantly constant as well:
\begin{equation}
D_\alpha \,\,\gamma _\beta = \partial _\alpha \,\gamma _\beta
-\gamma _\gamma\, \Gamma _{\alpha \beta }^\gamma
-\frac i4\left[ B_\alpha ^{mn}\sigma_{mn},\,\gamma _\beta \right] =0,
\label{Dgamma}
\end{equation}
where $D_\alpha $ is the full covariant derivative. It follows from (\ref
{Dgamma}) that
\begin{equation}
\left[ D_\alpha ,\,D_\beta \right] \,\gamma _\gamma =-i\,\left[
\hat{G}_{\alpha \beta },\,\gamma _\gamma \right]
-\gamma _\delta \,R_{\,\gamma
\,\alpha \beta }^\delta =0.  \label{G-R}
\end{equation}
Here $R_{\,\gamma \,\alpha \beta }^\delta $ is Riemann curvature tensor and
$\hat{G}_{\alpha \beta }$ is the commutator of covariant derivatives with
respect to the spin connection $B_\alpha ^{mn}$:
\begin{equation}
\hat{G}_{\alpha \beta }=i\,\left[ \partial _\alpha -\frac i4B_\alpha
^{mn}\sigma _{mn},\,\partial _\beta -\frac i4B_\beta ^{kl}\sigma
_{kl}\right] =\left(
\begin{array}{cc}
\hat{F}_{\alpha \beta }^{-} & 0 \\
0 & \hat{F}_{\alpha \beta }^{+}
\end{array}
\right) ;
\end{equation}
It is not a problem to rewrite the second term in (\ref{G-R}) as $\gamma
_\delta R_{\;\gamma \,\alpha \beta }^\delta \,=\frac i4R_{\quad \alpha \beta
}^{\delta \zeta }\left[ \sigma _{\delta \zeta ,\,}\gamma _\gamma \right] $.

Separating diagonal blocks of the matrices $\hat{G}_{\alpha \beta }$ and
$\sigma _{\delta \zeta }$ one obtains the relation between the compensating
field strength and the Riemann curvature of the space:
\begin{mathletters}
\label{F=R/4}
\begin{eqnarray}
\hat{F}_{\alpha \beta }^{+} &=&-\frac 14R_{\,\alpha \beta }^{\quad \gamma
\delta }\hat{\xi}_{\gamma \delta }^{+};  \label{F=R/4a} \\
\hat{F}_{\alpha \beta }^{-} &=&-\frac 14R_{\,\alpha \beta }^{\quad \gamma
\delta }\hat{\xi}_{\gamma \delta }^{-}.  \label{F=R/4b}
\end{eqnarray}
\end{mathletters}
Thus $\hat{F}_{\alpha \beta }^{+}=\hat{F}_{\alpha \beta }^{-}=0$ provided
that $R_{\,\alpha \beta }^{\quad \gamma \delta }=0$. Simple changes of
variables $x^\mu \rightarrow q$ $^\alpha $ do not generate curvature and
both compensating fields are pure gauges.

\subsection{Relation between $\hat{\xi}_{ab}$ and $\hat{\eta}_{\mu \nu }$
symbols\label{xi<->eta}}

As soon as the vector-potentials $\hat{A}_\alpha ^{+}$ and $\hat{A}_\alpha
^{-}$ are pure gauges they may be represented as
\begin{equation}
\hat{A}_\alpha ^{+}(q)=i\,\Omega _{+}^{-1}(q)\,\partial _\alpha \Omega
_{+}(q)\qquad {\rm or}\qquad \hat{A}_\alpha ^{-}(q)=i\,\Omega
_{-}^{-1}(q)\,\partial _\alpha \Omega _{-}(q).  \label{A=OmDOm}
\end{equation}
This defines the $2\times 2$ matrices $\Omega _{+}$ and $\Omega _{-}$ up to
the left multiplication by a constant nondegenerate matrix $U$: $\Omega
\rightarrow U\,\Omega $. However the freedom may be eliminated by requiring
that $\Omega $ gauge rotated $\hat{\eta}_{\mu \nu }$ into $\hat{\xi}_{\alpha
\beta }$.

Let us rewrite the condition of $\hat{\xi}_{ab}$ being covariantly constant,
(\ref{Dxi^pm}), in Cartesian coordinates. It will read (presently we may
drop the $\pm $-superscripts):
\begin{equation}
D_\lambda \hat{\xi}_{\mu \nu }(x)=D_\lambda \;
\frac{\partial q^\alpha }{\partial x^\mu }
\frac{\partial q^\beta }{\partial x^\nu }\,e_\alpha
^a\,e_\beta ^b\,\hat{\xi}_{ab}=\partial _\lambda e_\mu ^a\,e_\nu ^b\,
\hat{\xi}_{ab}+e_\mu ^a\,e_\nu ^b\,\left[ \Omega ^{-1}\,
\partial _\lambda \Omega ,\,\hat{\xi}_{ab}\right] =0,
\end{equation}
that is equivalent to
\begin{equation}
\Omega ^{-1}\,\left[ \partial _\lambda \left( \Omega \,e_\mu ^a\,e_\nu ^b\,
\hat{\xi}_{ab}\Omega ^{-1}\right) \right] \Omega =0\quad \,{\rm or}\quad
\left( \Omega \,e_\mu ^a\,e_\nu ^b\,\hat{\xi}_{ab}\Omega ^{-1}\right) =
{\it const.}  \label{Om_xi_Om=c}
\end{equation}
We conclude that $\Omega \,e_\mu ^a\,e_\nu ^b\,\hat{\xi}_{ab}^{\pm }\Omega
^{-1}$ is a constant tensor that projects selfdual (or antiselfdual
respectively) antisymmetric tensors onto the $SU(2)$ group (see Appendix).
Hence it must be equal (up to a gauge rotation) to the corresponding
$\hat{\eta}_{\mu \nu }$. Uniform gauge rotations do not affect the
compensating fields (\ref{A=OmDOm}) and we may fix $\Omega $ up to a phase
factor $e^{i\alpha }$ by demanding that:
\begin{equation}
e_\mu ^a\,e_\nu ^b\,\hat{\xi}_{ab}=\Omega ^{-1}\,\hat{\eta}_{\mu \nu }\Omega
\qquad {\rm or}\qquad \Omega \,\hat{\xi}_{ab}=e_a^\mu \,e_b^\nu \,
\hat{\eta}_{\mu \nu }\Omega .  \label{Oxi=etaO}
\end{equation}

The question is whether there are solutions to these equations. One may
figure out the two conditions. First, the both sides of (\ref{Oxi=etaO})
must be normalized in the same way. In order to prove this we take the
square of the first equation:
\begin{equation}
e_\mu ^a\,e_\nu ^b\,\hat{\xi}_{ab}\,e^{c\,\mu }\,e^{d\,\nu }\,
\hat{\xi}_{cd}=
\Omega ^{-1}\,\hat{\eta}_{\mu \nu }\,\hat{\eta}^{\mu \nu }\,\Omega .
\end{equation}
This results into identity $\hat{\xi}_{ab}\,\hat{\xi}^{ab}=\hat{\eta}_{\mu
\nu }\,\hat{\eta}^{\mu \nu }=12$. Thus the first condition is fulfilled.

However this does not guarantee existence of solutions. It is necessary that
both sides of the equations were of same duality. This is the case if the
two sets $\tau _a$ and $\tau _\mu e_a^\mu $ differ by a rotation and/or an
even permutation.%
\footnote{Although gauge transformation rotate the traceless Pauli $\tau
$-matrices into each other, they do not affect the $\tau _4$-matrix. Hence
permutations can not be reduced to gauge transforms.}
(Another way is to say that the change of variables must respect parity.)
Gauge transformations do not interfere with duality and it suffice to
check the equivalence of (\ref {Oxi=etaO}) to their duals,
\begin{equation}
\frac 12\,\epsilon ^{\mu \nu \lambda \sigma }\,e_\lambda ^a\,e_\sigma ^b\,
\hat{\xi}_{ab}^{\pm }=\pm \Omega ^{-1}\,\hat{\eta}_{\mu \nu }^{\pm }\,\Omega
\qquad {\rm or}\qquad \Omega \,\hat{\xi}_{ab}^{\pm }=\pm \frac 12\,\epsilon
_{abcd}\,e^{\mu \,c}\,e^{\nu \,d}\,\hat{\eta}_{\mu \nu }^{\pm }\,\Omega .
\label{*Oxi=*etaO}
\end{equation}
A well-known example of a parity violating procedure is inversion
$q_\alpha =x_\alpha /x^2$ that may be implemented to derive the singular
gauge, (\ref {r/s_matrb}) from the regular one, (\ref{r/s_matra}). In
order to restore duality altered by the inversion the substitution $\eta
_{\mu \nu }\rightarrow -\bar{\eta}_{\mu \nu }$ is necessary.

\section{Instanton with the compensating field.}

\label{inst+comp}A naive way to transform an instanton to curvilinear
coordinates would be simply to change variables: $x^\mu \rightarrow q^\alpha
$ and $\hat{A}_\mu \rightarrow \hat{A}_\alpha =\hat{A}_\mu \,(\partial x^\mu
/\partial q^\alpha )$. It is more convenient however to accompany this by
one of the previously discussed gauge rotations. Let us begin with the
singular gauge, (\ref {r/s_matrb}), that contains the antiselfdual symbol
$\hat{\eta}_{\mu \nu }^{-}$.  This dictates the choice of the matrix
$\Omega _{-}$ for the transformation.  The combined change of variables
and gauge transform give:
\begin{equation}
\left. \hat{A}_\alpha ^\Omega \right| _{{\rm sing}}(q)=\Omega _{-}^{-1}(q)\,
\frac{\partial x^\mu }{\partial q^\alpha }\left. \hat{A}_\mu ^I
\right| _{{\rm sing}}(q)\,\Omega _{-}(q)+i\,
\Omega _{-}^{-1}(q)\frac{\partial \,\Omega
_{-}(q)}{\partial q_\alpha }\,,
\end{equation}
With the help of the relations (\ref{A=OmDOm}) and (\ref{Oxi=etaO}) this may
be immediately reduced to
\begin{equation}
\left. \hat{A}_\alpha ^\Omega \right| _{{\rm sing}}(q)=-\frac 12e_\alpha ^a\,
\hat{\xi}_{ab}^{-}\,e^{b\,\beta }\,\partial _\beta \ln \Pi _{{\rm sing}}(q)+
\hat{A}_\alpha ^{-}.  \label{A_s^I+A^-}
\end{equation}
On the other hand instanton field in the regular gauge depends on
$\hat{\eta}_{\mu \nu }^{+}$ and the matrix $\Omega _{+}$ must be employed:
\begin{equation}
\left. \hat{A}_\alpha ^\Omega \right| _{{\rm reg}}(q)=\frac 12e_\alpha ^a\,
\hat{\xi}_{ab}^{+}\,e^{b\,\beta }\,\partial _\beta \ln \Pi _{{\rm reg}}(q)+
\hat{A}_\alpha ^{+}.  \label{A_r^I+A^+}
\end{equation}

An interesting feature of the expressions (\ref{A_s^I+A^-}, \ref{A_r^I+A^+})
is that they do not contain $\Omega $ explicitly. One needs only
$\hat{A}_\alpha ^{\pm }$ which may be found directly from (\ref{A^pm}).
That is much easier than solving (\ref{Oxi=etaO}) for $\Omega $. From here
on we shall omit the superscript $\Omega $ in $\hat{A}_\alpha ^\Omega
(q)$. In fact if the sign of the topological charge is not crucial one may
apply the formulae (\ref{A_s^I+A^-}, \ref{A_r^I+A^+}) and (\ref{A^pm})
without checking duality properties of $\hat{\xi}_{ab}$ and
$\hat{\eta}_{\mu \nu }\,e_a^\mu \,e_b^\nu $.

The duality equation, (\ref{tp_chrg}a), in the non-Cartesian frame takes the
form
\begin{equation}
\hat{F}_{\alpha \beta }=\frac{\sqrt{g}}2\,\epsilon _{\alpha \beta \gamma
\delta }\,\hat{F}^{\gamma \delta }
\qquad {\rm or} \qquad
\hat{F}_{ab} = \frac12 \epsilon_{abcd}\,\hat{F}^{cd},  \label{F=*Fcurv}
\end{equation}
where $g=\det \left| \left| g_{\alpha \beta }\right| \right| $. The
topological charge is, (compare to (\ref{tp_chrg}b)),
\begin{equation}
q=\frac 1{32\pi ^2}\int \,\epsilon _{\alpha \beta \gamma \delta }\,{\rm tr\,}
\hat{F}^{\alpha \beta }\,\hat{F}^{\gamma \delta }\,d^4q=\frac 1{32\pi
^2}\int \,\epsilon _{abcd}\,{\rm tr\,}\hat{F}^{ab}\,\hat{F}^{cd}\,\sqrt{g}
d^4q.  \label{tp_chrg_crv}
\end{equation}

We shall calculate the field strength in two ways. First we shall derive a
general expression that works in multiinstanton case as well. Then we
shall demonstrate that for one pseudoparticle the formulae do simplify
both in regular and singular gauges.

First of all let us notice that we may keep the Levi-Civita connection
$\Gamma _{\alpha \beta }^\gamma $ in the covariant derivative since it
will drop out of the final result. For brevity we shall denote the first
addends in the right hand sides of (\ref{A_s^I+A^-}, \ref{A_r^I+A^+}) by
$\hat{A}_\alpha ^I$.
\begin{equation}
\hat{F}_{\alpha \beta }(q)=D_\alpha \left( \hat{A}_\beta ^I+\hat{A}_\beta
^{\pm }\right) -D_\beta \left( \hat{A}_\alpha ^I+\hat{A}_\alpha ^{\pm
}\right) +i\,\left[ \hat{A}_\alpha ^I+\hat{A}_\alpha ^{\pm },\,\hat{A}_\beta
^I+\hat{A}_\beta ^{\pm }\right] ,
\end{equation}
where $D_\alpha $ is the full covariant derivative:
\begin{equation}
D_\alpha \left( \hat{A}_\beta ^I+\hat{A}_\beta ^{\pm }\right) =\partial
_\alpha \left( \hat{A}_\beta ^I+\hat{A}_\beta ^{\pm }\right) -i\,\left[
\hat{A}_\alpha ^I+\hat{A}_\alpha ^{\pm },\,\hat{A}_\beta ^I+
\hat{A}_\beta ^{\pm
}\right] -\Gamma _{\alpha \beta }^\gamma \,\left( \hat{A}_\gamma ^I+
\hat{A}_\gamma ^{\pm }\right) .
\end{equation}
The compensating field $\hat{A}_\beta ^{\pm }$ is a pure gauge and does not
contribute to $\hat{F}_{\alpha \beta }$. Thus it's only role is to ensure
that the factor $e_\alpha ^a\,\hat{\xi}_{ab}^{\pm }\,e^{b\,\beta }$ is
covariantly constant. Hence covariant derivatives act only on the
logarithms:
\begin{eqnarray}
\left. \hat{F}_{\alpha \beta }^I\right| _{{\rm reg}} &=&\frac 12e_\beta ^a\,
\hat{\xi}_{ab}^{\pm }\,e^{b\,\gamma }\,D_\alpha \,\partial _\gamma \ln
\Pi_{{\rm reg}}-\frac 12e_\alpha ^a\,\hat{\xi}_{ab}^{\pm }\,e^{b\,\gamma
}\,D_\beta \,\partial _\gamma \ln \Pi _{{\rm reg}}-i\,\left[ \hat{A}_\alpha
^I,\,\hat{A}_\beta ^I\right] , \\
\left. \hat{F}_{\alpha \beta }^I\right| _{{\rm sing}} &=&-\frac 12e_\beta
^a\,\hat{\xi}_{ab}^{\mp }\,e^{b\,\gamma }\,D_\alpha \,\partial _\gamma \ln
\Pi _{{\rm sing}}+\frac 12e_\alpha ^a\,\hat{\xi}_{ab}^{\mp }\,e^{b\,\gamma
}\,D_\beta \,\partial _\gamma \ln \Pi _{{\rm sing}}-i\,\left[ \hat{A}_\alpha
^I,\,\hat{A}_\beta ^I\right],
\end{eqnarray}
and do not contain gauge terms:
\begin{equation}
D_\alpha \,\partial _\gamma \ln \Pi (q)=\partial _\alpha \,\partial _\gamma
\ln \Pi (q)-\Gamma _{\alpha \gamma }^\delta \,\partial _\delta \ln \Pi (q).
\end{equation}
The commutator can be calculated with the help of the identity
\begin{equation}
\left[ \hat{\xi}_{ab},\,\hat{\xi}_{cd}\right] =-2i\,\left( \pi _{bc}
\hat{\xi}_{ad}+\pi _{ad}\hat{\xi}_{bc}-\pi _{ac}\hat{\xi}_{bd}-
\pi _{bd}\hat{\xi}_{ac}\right) ,  \label{[xi,xi]b}
\end{equation}
that follows from the properties of the $SU(2)$ generators but may be proved
directly. Denoting for brevity $(\partial \Pi )^2=g^{\alpha \beta
}\,\partial _\alpha \Pi \,\partial _\beta \Pi $ we obtain:
\begin{eqnarray}
\left. \hat{F}_{\alpha \beta }^I\right| _{{\rm sing}} &=&\frac{(\partial \Pi
)^2}{2\Pi ^2}\left( g_{\alpha \gamma }-\frac{2\,\partial _\alpha \Pi
\,\partial _\gamma \Pi }{(\partial \Pi )^2}\right) e^{a\,\gamma }\,
\hat{\xi}_{ab}^{-}\,e^{b\,\delta }\left( g_{\delta \beta }-
\frac{2\,\partial _\delta
\Pi \,\partial _\beta \Pi }{(\partial \Pi )^2}\right) +  \nonumber \\
&&+\frac{D_\alpha \,\partial _\gamma \Pi }{2\Pi }e^{a\,\gamma }\,
\hat{\xi}_{ab}^{-}\,e_\beta ^b+e_\alpha ^{a\,}\,\hat{\xi}_{ab}^{-}\,
e^{b\,\delta }\frac{D_\beta \,\partial _\delta \Pi }{2\Pi }.
\label{F_sg/long}
\end{eqnarray}
Mind that so long we have not referred to the specific form of $\Pi (q)$
and to the instanton number in particular. Thus the formula works for the
multi-instanton solutions as well.

In the regular gauge (this automatically implies that the topological charge
is unity) the formula looks more compact (note that now
$\Pi = \Pi_{\rm reg}$):
\begin{equation}
\left. \hat{F}_{\alpha \beta }^I\right| _{{\rm reg}}=
\frac{(\partial \Pi )^2}{2\Pi ^2}
e_\alpha ^{a\,}\,\hat{\xi}_{ab}^{+}\,e_\beta ^{b\,}-\frac{D_\alpha
\,\partial _\gamma \Pi }{2\Pi }e^{a\,\gamma }\,\hat{\xi}_{ab}^{+}\,e_\beta
^b-e_\alpha ^{a\,}\,\hat{\xi}_{ab}^{+}\,e^{b\,\delta }\frac{D_\beta
\,\partial _\delta \Pi }{2\Pi }
\end{equation}
However it may be significantly simplified if we start from $\left.
\hat{F}_{\mu \nu }^I\right| _{{\rm reg}}$ in Cartesian coordinates,
\begin{equation}
\left. \hat{F}_{\mu \nu }^I\right| _{{\rm reg}}=-\frac{2\hat{\eta}_{\mu \nu
}^{+}}{\left( r^2+\rho ^2\right) ^2},  \label{F_mu_nu_reg}
\end{equation}
with $r^2=x_\mu ^2$ and transform it to $q$-coordinates. Obviously the
transformation affects only $\hat{\eta}_{\mu \nu }^{+}$. According to the
$\Omega _+$ version of (\ref{Oxi=etaO}) the result is:
\begin{equation}
\left. \hat{F}_{\alpha \beta }^I\right| _{{\rm reg}}=-\frac{2\,e_\alpha ^a\,
\hat{\xi}_{ab}^{+}\,e_\beta ^b}{\left( r^2+\rho ^2\right) ^2}
\qquad {\rm or}\qquad
\left. \hat{F}_{ab}^I\right| _{{\rm reg}}
=-\frac{2\,\hat{\xi}_{ab}^{+}}{\left( r^2+\rho ^2\right) ^2}
\label{F_1_reg}
\end{equation}
The last version is explicitly selfdual.

The expression (\ref{F_sg/long}) in the one instanton case can be simplified
too. Once more we shall start from (\ref{F_mu_nu_reg}). Remember that
Cartesian pseudoparticle in singular gauge can be obtained from that in the
regular gauge by means of the transformation (\ref{A_s=OA_rO+OdO}). This
results into $\left. \hat{F}_{\mu \nu }^I\right| _{{\rm sing}}=
\hat{N}_{+}^{-1}\left. \hat{F}_{\mu \nu }^I\right| _{{\rm
reg}}\hat{N}_{+}$. Now we should carry out the simultaneous change of
variables and $\Omega _{-}$ gauge transform:
\begin{equation}
\left. \hat{F}_{\alpha \beta }^I\right| _{{\rm sing}}=-\frac{\partial x^\mu
}{\partial q^\alpha }\frac{\partial x^\nu }{\partial q^\beta }\,\Omega
_{-}^{-1}\,\hat{N}_{+}^{-1}\frac{2\hat{\eta}_{\mu \nu }^{+}}{\left( r^2+\rho
^2\right) ^2}\hat{N}_{+}\,\Omega _{-}.
\end{equation}
It is convenient to rewrite $\hat{N}$-matrices as: $\hat{N}_{+}=\tau _\mu
^{\dagger }\,\partial _\mu r$ and $\hat{N}_{+}^{-1}=
\tau _\mu ^{\vphantom{\dagger}}\,\partial _\mu r$
(note that the Cartesian $g_{\mu \nu }=\delta _{\mu \nu }$). After a
little algebra with the use of (\ref {Oxi=etaO}) one obtains:
\begin{equation}
\left. \hat{F}_{\alpha \beta }^I\right| _{{\rm sing}}=-{\cal N}_{+}^{-1}
\frac{2\hat{\xi}_{\alpha \beta }^{+}}{\left( r^2+\rho ^2\right) ^2}
{\cal N}_{+};\quad {\cal N}_{+}
=\tau _a^{\dagger }\,e^{a\,\alpha }\partial _\alpha
r;\quad {\cal N}_{+}^{-1}=\tau _a^{\vphantom{\dagger}}\,e^{a\,\alpha
}\partial _\alpha r.  \label{F_1_sing}
\end{equation}

We note that all connections have dropped out giving the compact results
(\ref{F_1_reg}, \ref{F_1_sing}). Unfortunately for multipseudoparticle
solutions one still has to apply the clumsy formula (\ref{F_sg/long}). Later
in the Sect.~\ref{Sect_O(4)-inst} we shall describe the case when matrices
${\cal N}_{+}$ become degenerate ({\em i.~e.\/} ${\cal N}_{+}=-i$) so that
(\ref{F_1_reg}) and (\ref{F_1_sing}) coincide.

It is obvious that all the reasoning may be literally repeated for
anti-instantons. The only difference is that one has to interchange
$\hat{\xi}_{ab}^{+}\longleftrightarrow $ $\hat{\xi}_{ab}^{-}$ and
$\tau _a^{\vphantom{\dagger}}\longleftrightarrow \tau _a^{\dagger }$.
The matrices ${\cal N}_{+}$, ${\cal N}_{+}^{-1}$ must be substituted by
${\cal N}_{-} = {\cal N}_{+}^{-1}$ and ${\cal N}_{-}^{-1}={\cal N}_{+}$.
Certainly the signs of the right hand sides of the equations
(\ref{F=*Fcurv}) must be reversed as well.

\section{Instanton in $O(4)$-spherical coordinates.}

\label{Sect_O(4)}In order to show how the general theory works we shall
apply it to the familiar case. Let us derive explicit formulae for one
instanton placed at the origin of the 4-dimensional spherical coordinates.
The example happens to be instructive and reveals two unexpected features.
First, it turns out that (in this particular setting) our prescription
converts pseudoparticles both in singular and regular gauges to the same
form. In a sense this is an evidence in favor of the approach. Second, the
compensating vector potential exhibits a singularity. From the first sight
the singularity of the field that is a pure gauge is of no importance.
However in presence of a pseudoparticle the singularity comes to life and
contributes to the Chern-Simons number.

We begin from a general description of the 4-dimensional spherical
coordinates and then focus on the pseudoparticles. Finally we present the
calculation of the Chern-Simons number.

\subsection{4-dimensional spherical coordinates.}

Spherical coordinates make a natural choice for problems involving single
euclidean pseudoparticle. The set of spherical coordinates includes radius
and three angles: $q^\alpha =(r,\,\theta ,\,\phi ,\,\chi )$. The polar axis
is aligned with $x^1$ and
\begin{equation}
x^1=r\cos \chi ;\quad x^2=r\sin \chi \sin \theta \cos \phi ;\quad x^3=r\sin
\chi \sin \theta \sin \phi ;\quad x^4=r\sin \chi \cos \theta .
\end{equation}
The metric tensor and the vierbein are:
\begin{equation}
g_{\alpha \beta }={\rm diag}(1,\,r^2\sin ^2\chi ,\,r^2\sin ^2\chi \sin
^2\theta ,\,r^2);\quad e_\alpha ^a={\rm diag}(1,\,r\sin \chi ,\,r\sin \chi
\sin \theta ,\,r)\,.
\end{equation}
It is convenient to introduce the matrix notation for the Levi-Civita
symbols, $\hat{\Gamma}_\alpha =\left| \left| \Gamma _{\alpha \beta }^\gamma
\right| \right| $. The standard calculation gives:
\begin{eqnarray}
\hat{\Gamma}_\theta &=&\left(
\begin{array}{cccc}
0 & -r\,{\sin ^2\chi } & 0 & 0 \\
{\frac 1r} & 0 & 0 & \cot \chi \\
0 & 0 & \cot \theta & 0 \\
0 & -\frac 12\sin 2\chi & 0 & 0
\end{array}
\right) ;\quad \hat{\Gamma}_\chi =\left(
\begin{array}{cccc}
0 & 0 & 0 & -r \\
0 & \cot \chi & 0 & 0 \\
0 & 0 & \cot \chi & 0 \\
{\frac 1r} & 0 & 0 & 0
\end{array}
\right) ; \\
\hat{\Gamma}_\phi &=&\left(
\begin{array}{cccc}
0 & 0 & -r\,{\sin ^2\chi }\,{\sin ^2\theta } & 0 \\
\ 0 & 0 & -\frac 12\sin 2\theta & \ 0 \\
\ {\frac 1r} & \cot \theta & 0 & \cot \chi \\
\ 0 & 0 & -\frac 12\sin 2\chi \,{\sin ^2\theta } & 0
\end{array}
\right) ;\quad \hat{\Gamma}_r=\left(
\begin{array}{cccc}
0 & 0 & 0 & 0 \\
0 & {\frac 1r} & 0 & 0 \\
0 & 0 & {\frac 1r} & 0 \\
0 & 0 & 0 & {\frac 1r}
\end{array}
\right) ; \nonumber
\end{eqnarray}

By means of the relation (\ref{R_alpha}) we may find the spin connection.
Sticking once again to the matrix notation, $\hat{R}_\alpha =\left| \left|
R_{\alpha \,b}^a\right| \right| $, one obtains:
\begin{eqnarray}
\hat{R}_\theta &=&\left(
\begin{array}{cccc}
0 & -\sin \chi & 0 & 0 \\
\sin \chi & 0 & 0 & \cos \chi \\
0 & 0 & 0 & 0 \\
0 & -\cos \chi & 0 & 0
\end{array}
\right) ;\quad \hat{R}_\chi =\left(
\begin{array}{cccc}
0 & 0 & 0 & -1 \\
0 & 0 & 0 & 0 \\
0 & 0 & 0 & 0 \\
1 & 0 & 0 & 0
\end{array}
\right) ; \\
\hat{R}_\phi &=&\left(
\begin{array}{cccc}
0 & 0 & -\sin \chi \sin \theta & 0 \\
0 & 0 & -\cos \theta & 0 \\
\sin \chi \sin \theta & \cos \theta & 0 & \cos \chi \sin \theta \\
0 & 0 & -\cos \chi \sin \theta & 0
\end{array}
\right) ;\quad \hat{R}_r=0.  \nonumber
\end{eqnarray}

The compensating gauge potentials $\hat{A}^{+}$ and $\hat{A}^{-}$ depend on
the particular choice of $\tau $-matrices. Let us remind that the
antisymmetrized products $\hat{\xi}_{ab}^{+}$ and $\hat{\xi}_{ab}^{-}$ must
be of correct duality so that the equations (\ref{Oxi=etaO}) and (\ref
{*Oxi=*etaO}) were equivalent. For example, if $\tau _x$, $\tau _y$, $\tau
_z $ stand for standard Pauli matrices we can take:
\begin{equation}
\tau _a=(i,\;\tau _z,\,\tau _y,\,\tau _x);\qquad \tau _a^{\dagger
}=(-i,\;\tau _z,\,\tau _y,\,\tau _x).  \label{tau2+2}
\end{equation}
Convolutions of the $\tau $-matrices (\ref{tau2+2}) with the spin
connection, (\ref{A^pm}), result into the compensating gauge fields:
\begin{eqnarray}
\hat{A}_\theta ^{\pm } &=&-\frac{\tau _y}2{cos\chi \mp }\frac{\tau _z}2{\sin
\chi ;\qquad }\hat{A}_\chi ^{\pm }={\mp }\frac{\tau _x}2;  \label{A^pmO(4)}
\\
\hat{A}_\phi ^{\pm } &=&-\frac{\tau _x}2\cos \theta {\mp }
\frac{\tau _y}2\sin \chi \sin \theta +
\frac{\tau _z}2\cos \chi \sin \theta ;{\qquad }\hat{A}_r^{\pm }=0;
\nonumber
\end{eqnarray}
Either of these fields is a pure gauge generated by the corresponding
unitary ($\Omega _{\pm }^{-1}=\Omega _{\pm }^{\dagger }$) matrix:
\begin{equation}
\Omega _{\pm }=\left(
\begin{array}{cc}
\sin \frac \theta 2\sin \frac{\phi \mp \chi }2+i\cos \frac \theta 2\cos
\frac{\phi \pm \chi }2 & -\cos \frac \theta 2\sin
\frac{\phi \pm \chi }2+i\sin \frac \theta 2\cos \frac{\phi \mp \chi }2 \\
\cos \frac \theta 2\sin \frac{\phi \pm \chi }2+i\sin \frac \theta 2\cos
\frac{\phi \mp \chi }2 & \hphantom{-}\sin \frac \theta 2\sin \frac{\phi \mp
\chi }2-i\cos \frac \theta 2\cos \frac{\phi \pm \chi }2
\end{array}
\right) .
\end{equation}
The matrices $\Omega _{+}$ and $\Omega _{-}$ for selfdual and antiselfdual
cases differ by the sign of the polar angle $\chi $.

The compensating connection is singular since neither $\hat{A}_\theta ^{\pm
} $ nor $\hat{A}_\phi ^{\pm }$ go to zero near polar axes $\chi =0$ and
$\theta =0$. As long as the field is a pure gauge this singularity is not
observable. However as we shall see below in presence of physical fields
it may tell on gauge variant quantities.

\subsection{Instantons in 4-spherical coordinates.}

\label{Sect_O(4)-inst}We shall consider instantons in singular and regular
gauges with their centers at the origin. The analysis reveals an amusing
coincidence. Remember that instanton gauge potentials in these two cases
involve different 'tHooft symbols $\hat{\xi}_{ab}^{+}$ and
$\hat{\xi}_{ab}^{-}$. Hence according to our prescription the compensating
vector potentials are $\hat{A}^{+}$ and $\hat{A}^{-}$ respectively. It
turns out that if the $\tau $-matrices are taken in the form
(\ref{tau2+2}) then equation (\ref{A_s^I+A^-}) for the singular gauge and
that (\ref{A_r^I+A^+}) for the regular one give the same result. Let us
take $\Pi _{{\rm reg}}$ and $\Pi _{{\rm sing}}$ (\ref{r/s_matr}) in the
form:
\begin{equation}
\Pi _{{\rm reg}}(r)=r^2+\rho ^2;
\qquad {\rm and}\qquad
\Pi _{{\rm sing}}(r)=1+\rho ^2/r^2.
\end{equation}
Substituting those into the equations (\ref{A_s^I+A^-}, \ref{A_r^I+A^+})
with the compensating potentials given by (\ref{A^pmO(4)}) we obtain:
\begin{mathletters}
\label{instO(4)}
\begin{eqnarray}
{\cal A}_r^{\pm } &=&0;  \label{instO(4)a} \\
{\cal A}_\theta ^{\pm } &=&-\frac{\tau _y}2{cos\chi \pm }\frac{\tau _z}2
\sin \chi \left( 1-\frac{2\rho ^2}{r^2+\rho ^2}\right) ;
\label{instO(4)b} \\
{\cal A}_\phi ^{\pm } &=&-\frac{\tau _x}2\cos \theta {\pm }
\frac{\tau _y}2\sin \chi \sin \theta
\left( 1-\frac{2\rho ^2}{r^2+\rho ^2}\right) +
\frac{\tau _z}2\cos \chi \sin \theta ;  \label{instO(4)c} \\
{\cal A}_\chi ^{\pm } &=&{\pm }\frac{\tau _x}2
\left( 1-\frac{2\rho ^2}{r^2+\rho ^2}\right) ;  \label{instO(4)d}
\end{eqnarray}
\end{mathletters}
Note that the potentials ${\cal A}^{\pm }$ interpolate between
$\hat{A}^{\pm }$ at the origin and $\hat{A}^{\mp }$ at infinity:
\begin{equation}
\lim_{r\rightarrow 0}{\cal A}_\alpha ^{\pm }(q)=
\hat{A}_\alpha ^{\pm}(q);
\qquad {\rm and}\qquad
\lim_{r\rightarrow \infty }{\cal A}_\alpha ^{\pm}(q)
=\hat{A}_\alpha ^{\mp }(q).  \label{+interp-}
\end{equation}

The field strength is given by the expression (\ref{F_1_reg}). The
comparison of the latter with (\ref{F_1_sing}) proves that as a rule regular
and singular gauges are different. However in present case the matrices
${\cal N}_{\pm }=\mp i$ and the transform is trivial. The degeneracy is
specific to our choice of coordinates. Note that we took $q^1=r$ and
associated with it $\tau _1=i$. According to (\ref{F_1_sing}) these are
necessary and sufficient conditions of the coincidence of the two gauges in
curvilinear coordinates. Hence this coincidence is stable with respect to
reparametrizations of the angles $\theta $, $\phi $, $\chi $ and
$\tau $-matrices as long as $r$ and $\tau _1$ stay intact.

\subsection{Singularity and the Chern-Simons number.}

It is well known that the topological charge density, (\ref{tp_chrg_crv}),
may be represented as a total divergence:
\begin{equation}
\frac{\epsilon _{\alpha \beta \gamma \delta }\,{\rm tr\,}\hat{F}^{\alpha
\beta }\,\hat{F}^{\gamma \delta }}{32\pi ^2}\,=\partial _\alpha K^\alpha
=\partial _\alpha \frac{\epsilon ^{\alpha \beta \gamma \delta }}{16\pi ^2}
{\,\rm tr\,}\left( \hat{A}_\beta \,\hat{F}_{\gamma \delta }
+\frac{2i}3\hat{A}_\beta \hat{A}_\gamma \hat{A}_\delta \right) .
\end{equation}
According to the Gauss theorem the space integral of the {\em LHS\/} may be
reduced to the surface integral of $K^\alpha $ over the boundary. However
the current $K^\alpha $ is not gauge invariant. Thus the distribution of
$K^\alpha $ over the boundary depends on gauge even though the topological
charge $q=\oint K^\alpha \,dS_\alpha $ does not. A text book example is an
instanton in the $\hat{A}_4=0$ gauge. Here the two nonzero contributions to
$q$ come from the hyperplanes $x_4=\pm \infty $, ($i,\,j,\,k=1,\,2,\,3$):
\begin{equation}
q(\hat{A}_4=0)=\oint K^\alpha \,dS_\alpha =
-\left. \int \frac{\,d^3\vec{x}}{16\pi ^2}
\epsilon ^{ijk}\,{\rm tr\,}\left( \hat{A}_i\,\hat{F}_{jk} + \frac{2i}3
\hat{A}_i\hat{A}_j\hat{A}_k\right) \right|_{x_4=-\infty }^{x_4=+\infty}.
\label{q=Q_0-Q_0}
\end{equation}
The integral $Q_0(x_4)=\int d^3\vec{x}\,K^4(x_4)$ over a 3-dimensional
manifold is called the Chern-Simons number. In the example above $Q_0(\pm
\infty )=\pm \frac 12$ leading to $q=1$. Thus the topological charge may
be interpreted as the change of the Chern-Simons number.

Let us try to carry out the same procedure for the gauge field (\ref
{instO(4)}) in spherical coordinates. At the first sight there is a striking
similarity with the $\hat{A}_4=0$ case. According to (\ref{+interp-}) now
${\cal A}_r=0$ and the instanton interpolates between $\hat{A}^{+}$ at $r=0$
and $\hat{A}^{-}$ at $r=\infty $. It would be nice to take the radius (or
rather $\rho \ln r/\rho$) for the 4-th coordinate and to associate the
Chern-Simons number $Q(r)=\oint_{S_r^3}K^r(r)\,d\theta \,d\chi \,d\phi $
with the sphere $ S_r^3$ of the radius $r$. The analogy would be complete
provided that one had the contributions $Q(\infty )=\frac 12$ from the
infinite sphere and $Q(0)=-\frac 12$ from the infinitesimal sphere at
the origin%
\footnote{In regular gauge, (\ref{r/s_matra}), the Chern-Simons number is
concentrated at infinity $q=Q_r(\infty )=1$, in contrast to the singular
one, (\ref {r/s_matrb}), where $q=-Q_s(0)=1$.}.  However the explicit
calculation proves that this is not the case because the singular line
$\theta =0$ carries the Chern-Simons number as well.

The topological charge in 4-spherical coordinates is given by the integral
over a half infinite parallelepiped $\{0\leq r\leq \infty ,\,0\leq \chi \leq
\pi ,\,0\leq \theta \leq \pi ,\,0\leq \phi \leq 2\pi \}$:
\begin{equation}
q=\frac 1{32\pi ^2}\int_0^\infty dr\int_0^\pi d\chi \,\int_0^\pi d\theta
\,\int_0^{2\pi }d\phi \,\epsilon _{\alpha \beta \gamma \delta }\,{\rm tr}\,
\hat{F}^{\alpha \beta }\,\hat{F}^{\gamma \delta }.
\label{q_spher}
\end{equation}
When applying the Gauss theorem one must
take into account the entire boundary including lateral faces. The
components of the Chern-Simons current are:
\begin{mathletters}
\label{K^alph}
\begin{eqnarray}
K^r &=&\frac{\sin \theta }{2\pi ^2}
\left( 1-\frac{2\rho ^2}{r^2+\rho ^2}\right)
\left( \frac 18+
\frac{\rho ^2\,r^2\,\sin ^2\chi }{\left( r^2+\rho^2\right) ^2}\right) ;
\label{K^alpha} \\
K^\chi &=&-\frac{\rho ^2\,r\,\sin 2\chi \,\sin \theta }{4\pi ^2\,\left(
r^2+\rho ^2\right) ^2};  \label{K^alphb} \\
K^\theta &=&-\frac{\rho ^2\,r\,\cos \theta }{4\pi ^2\,\left( r^2+\rho
^2\right) ^2};\qquad K^\phi =0.  \label{K^alphc}
\end{eqnarray}
\end{mathletters}

Converting (\ref{q_spher}) to the surface integral and keeping only nonzero
pieces we write:
\begin{equation}
q=\left. \int_0^\pi d\chi \,\int_0^\pi d\theta \,\int_0^{2\pi }d\phi
\,K^r\right| _{r=0}^{r=\infty }+\left. \int_0^\infty dr\,\int_0^\pi d\chi
\,\,\int_0^{2\pi }d\phi \,K^\theta \right| _{\theta =0}^{\theta =\pi }.
\label{q=4Q}
\end{equation}
The first addend in the {\em RHS\/} is $Q(\infty )-Q(0)$ and the second
comes from the integration along the ``hyperstrips'' $\theta =0,\,\pi $.
Each of the four addends contributes $\frac 14$ to the final value $q=1$. We
see that the singularity of the compensating gauge field is not absolutely
harmless. The reason for it's coming to being was that $K^\alpha $ is not a
gauge invariant.

It is easy to see that $K^\theta = \epsilon ^{\theta \phi r \chi}
{\,\rm tr\,}\hat{A}_\phi \,\hat{F}_{r\chi }/8\pi^2$. Multiplication of
the singular gauge potential $\hat{A}_\phi $ by the instanton field
$\hat{F}_{r\chi }$ results into the singularity of $K^\theta $. Thus the
singularity of the seemingly unobservable gauge field $\hat{A}^{\pm }$
leads to nonzero consequences in presence of the physical field
$\hat{A}^I$. In general singular gauge transformations may generate
singularities of gauge variant quantities. Obviously this is specific
neither to the instanton nor to the current choice of the $O(4)$-spherical
coordinates.

\section{Cylindrical coordinates}

\label{cylinder}Let us describe two other coordinate systems that are
relevant to physics. We shall start from $2+2$ cylindrical coordinates,
{\em i.~e.\/} the geometry of the thick euclidean QCD-string or vortex.
Then we shall consider $3+1$ cylindrical coordinates which are
characteristic for the bag model or a glueball.

\subsection{2+2 cylindrical coordinates.}

These coordinates may be appropriate for objects with axial symmetry, such
as strings, vortices or quark-antiquark pairs. We parametrize the
$x^1x^2$-plane by polar coordinates $q^1=r$ and $q^2=\phi $ and leave
$q^{3,\,4}=x^{3,\,4}$:
\begin{equation}
x^1=r\cos \phi ;\quad x^2=r\sin \phi ;\quad x^3=z;\quad x^4=t.
\end{equation}
The metric tensor and the vierbein are respectively:
\begin{equation}
g_{\alpha \beta }={\rm diag\,}(1,\,r^2,\,1,\,1)\qquad {\rm and}\qquad
e_\alpha ^a={\rm diag\,}(1,\,r,\,1,\,1).
\end{equation}
Only three of the Levi-Civita symbols, (\ref{Gamma}), are not zero:
\begin{equation}
\Gamma _{\phi \phi }^r=-r\qquad {\rm and}\qquad \Gamma _{r\phi }^\phi
=\Gamma _{\phi r}^\phi =r^{-1}.
\end{equation}
The direct calculation (see equation (\ref{R_alpha})) proves that the spin
connection has only one nonzero component. Namely:
\begin{equation}
\hat{R}_\phi =\left(
\begin{array}{cccc}
0 & -1 & 0 & 0 \\
1 & 0 & 0 & 0 \\
0 & 0 & 0 & 0 \\
0 & 0 & 0 & 0
\end{array}
\right) ;\qquad \hat{R}_r=\hat{R}_z=\hat{R}_t=0.
\end{equation}
Convolutions of the $\tau $-matrices (\ref{tau2+2}) with the spin connection
$R$, (\ref{A^pm}), give the compensating fields that are singular at $r=0$:
\begin{equation}
\hat{A}_\phi ^{\pm }={\mp }\frac{\tau _z}2;
\qquad
\hat{A}_r^{\pm }=\hat{A}_z^{\pm }=\hat{A}_t^{\pm }=0.
\end{equation}

Finally, for completeness we present explicitly the unitary gauge matrices
$\Omega _{+}$ and $\Omega _{-}$.
\begin{equation}
\Omega _{\pm }=\frac 1{\sqrt{2}}\left(
\begin{array}{cc}
\exp \pm \frac{i\phi }2 & -\exp \mp \frac{i\phi }2 \\
\exp \pm \frac{i\phi }2 & \hphantom{-}\exp \mp \frac{i\phi }2
\end{array}
\right) ;\qquad \Omega _{\pm }^{-1}=\Omega _{\pm }^{\dagger };  \label{Om2+2}
\end{equation}
The difference between the matrices $\Omega _{+}$ and $\Omega _{-}$ is in
the sign of the polar angle $\phi $ (corresponding to the `left' and `right'
rotations).

\subsection{3+1 cylindrical coordinates.}

This geometry is characteristic for objects that are spherically symmetrical
in 3-di\-men\-sions. In addition to the famous MIT-bag and the glueball
one may list the monopole and the caloron that is a periodic instanton
chain along the 4th axis.  It takes the place of the instanton in thermal
problems, \cite{HS}. Now we parametrize the ``spatial'' sector, {\em
i.~e.\/} $x^1$, $x^2$, $x^3$ by spherical coordinates $q^1=r$, $q^2=\theta
$ and $q^3=\phi $ leaving $q^4=x^4$.
\begin{equation}
x^1=r\sin \theta \cos \phi ;\quad x^2=r\sin \theta \sin \phi ;\quad
x^3=r\cos \theta ;\quad x^4=t.  \label{3+1cyl}
\end{equation}
The metric tensor and the vierbein are well known:
\begin{equation}
g_{\alpha \beta }={\rm diag}(1,\,r^2,\,r^2\sin ^2\theta ,\,1);\qquad
e_\alpha ^a={\rm diag}(1,\,r,\,r\sin \theta ,\,1).
\end{equation}

Since the transformation (\ref{3+1cyl}) affects only spatial sector the
$t$-components, $\hat{\Gamma}_t=0$, and $\Gamma $'s are essentially
3-dimensional:
\begin{equation}
\hat{\Gamma}_r=\left(
\begin{array}{ccc}
0 & 0 & 0 \\
0 & \frac 1r & 0 \\
0 & 0 & \frac 1r
\end{array}
\right) ;\quad \hat{\Gamma}_\theta =\left(
\begin{array}{ccc}
0 & -r & 0 \\
\frac 1r & 0 & 0 \\
0 & 0 & \cot \theta
\end{array}
\right) ;\quad \hat{\Gamma}_\phi =\left(
\begin{array}{ccc}
0 & 0 & -r\sin ^2\theta \\
0 & 0 & -\frac 12\sin 2\theta \\
\frac 1r & \cot \theta & 0
\end{array}
\right) .
\end{equation}
The nonzero components of the spin connection are:
\begin{equation}
\hat{R}_\theta =\left(
\begin{array}{cccc}
0 & -1 & 0 & 0 \\
1 & 0 & 0 & 0 \\
0 & 0 & 0 & 0 \\
0 & 0 & 0 & 0
\end{array}
\right) ;\quad \hat{R}_\phi =\left(
\begin{array}{cccc}
0 & 0 & -\sin \theta & 0 \\
0 & 0 & -\cos \theta & 0 \\
\sin \theta & \cos \theta & 0 & 0 \\
0 & 0 & 0 & 0
\end{array}
\right) ;\quad \hat{R}_r=\hat{R}_t=0.
\end{equation}

We shall use the same set of $\tau $-matrices as before, (\ref{tau2+2}).
This leads to the following compensating connections:
\begin{equation}
\hat{A}_\theta ^{\pm }={\mp }\frac{\tau _z}2;
\quad \hat{A}_\phi ^{\pm }=
-\frac{\tau _x}2\cos \theta {\mp }\frac{\tau _y}2\sin \theta ;
\quad \hat{A}_r^{\pm }=\hat{A}_t^{\pm }=0.  \label{A3+1}
\end{equation}
This vector potential is singular at $\theta =0$. The gauge
matrices generating the above connections are:
\begin{equation}
\Omega _{\pm }\frac 1{\sqrt{2}}\left(
\begin{array}{cc}
\exp \frac i2\left( \mp \theta -\phi -\frac \pi 2\right) & -\exp \frac
i2\left( \pm \theta -\phi -\frac \pi 2\right) \\
\exp \frac i2\left( \mp \theta +\phi +\frac \pi 2\right) & \hphantom{-}\exp
\frac i2\left( \pm \theta +\phi +\frac \pi 2\right)
\end{array}
\right) ;\quad \Omega _{\pm }^{-1}=\Omega _{\pm }^{\dagger }  \label{Om3+1}
\end{equation}
We note again that $\Omega _{+}$ and $\Omega _{-}$ differ by the sign of the
phase $\theta $.

\section*{Conclusion}

Our purpose was to analyze exact multiinstanton solutions in non-Cartesian
coordinates. We showed that 'tHooft's $5N$-parametric {\em Ansatz\/} can
be economically generalized to curvilinear coordinates. The $\hat{\eta}_{\mu
\nu }$-symbols with coordinate indices are replaced by the
$\hat{\xi}_{ab}$-symbols with the vierbein ones. The price for
$\hat{\xi}_{ab}$ being constant is the appearance of the compensating
gauge field. The origin of the latter is the coordinate dependent gauge
transformation. Thus the proposed solution is a gauge rotated version of
the original 'tHooft's {\em Ansatz.}

The compensating gauge potential may be obtained in a straightforward
manner. The calculation proceeds in three steps.
\begin{enumerate}
\item  Starting from the metrics $g_{\alpha \beta }$ one may find the
Levi-Civita connections $\Gamma _{\beta \gamma }^\alpha $, (\ref{Gamma}).

\item  Covariant differentiation of the vierbein $e_a^\alpha $ leads to the
spin connection $R_{\alpha \,\beta }^a$, (\ref{R_alpha}).

\item  Finally the convolution of the spin connection with $\tau $-matrices
gives the compensating gauge potential, (\ref{A^pm}).
\end{enumerate}

The first two points are the standard calculation of the spin connection.
The last one is nothing but projecting the selfdual (or antiselfdual)
component of the antisymmetric tensor $R_\alpha ^{ab}$ onto the $SU(2)$
gauge group.

The gauge potential of a pseudoparticle is a sum of the instanton part that
is similar to the 'tHooft's formula and the compensating field, (\ref
{A_s^I+A^-}, \ref{A_r^I+A^+}). Having constant $\hat{\xi}_{ab}$,
(\ref{xi_ab}), and covariantly constant
$\hat{\xi}_{\alpha \beta }$-symbols, (\ref {xi_alphabet}), notably
simplifies calculations and is worth the appearance of the additive
compensating background.  Singularities of the compensating connection do
not spoil physical quantities.

I would like to thank A.~A.~Rosly and E.~Gozzi for encouraging
discussions. It is a pleasure to acknowledge the hospitality of the
Department of Theoretical Physics of Trieste University that I enjoyed
several times and where this research was completed as well as financial
support of INFN and COFIN-97-MURST of Italy. The work was done with
partial support of the RFBR grant 97-02-16131.
				   
\section*{Appendix.~~~Duality properties of the
$\hat{\xi}_{\mu\nu}$-symbols.} 

It was mentioned in Section~\ref{xi<->eta} that $\Omega \,e_\mu ^a\,e_\nu
^b\,\hat{\xi}_{ab}^{+}\,\Omega ^{-1}$ is a constant selfdual (antiselfdual
for $\hat{\xi}_{ab}^{-}$) tensor, see (\ref{Om_xi_Om=c}). We shall prove the
second part of this statement. First let us remind how coordinate changes
$x^\mu \rightarrow q^\alpha $ affect the Levi-Civita antisymmetric
pseudotensors. The change of variables leads to the substitution:
\begin{equation}
\epsilon ^{\mu \nu \lambda \sigma }
\rightarrow E^{\alpha \beta \gamma \delta}
=\frac{\partial q^\alpha }{\partial x^\mu }
\frac{\partial q^\beta }{\partial x^\nu }
\frac{\partial q^\gamma }{\partial x^\lambda }
\frac{\partial q^\delta }{\partial x^\sigma }
\epsilon ^{\mu \nu \lambda \sigma }
=\det \left| \left| \frac{\partial q^\alpha }{\partial x^\mu }\right| \right|
\epsilon ^{\alpha \beta \gamma \delta }
=(\pm )\frac{\epsilon ^{\alpha \beta \gamma \delta }}{\sqrt{g}},
\label{e^mu->e^alpha}
\end{equation}
where $g=\left| \det \left| \left| g_{\alpha \beta }\right| \right| \right| $
and $\epsilon ^{\alpha \beta \gamma \delta }$ is the ordinary
$\epsilon $-symbol ($\epsilon ^{1234}=1$). The sign of the last expression
depends on the relative orientation of the $q$ and $x$ coordinate systems.
We shall assume that the sign is plus.

The second formula relates $\epsilon $-symbol with coordinate (greek) and
vierbein (latin) indices:
\begin{equation}
\epsilon ^{\alpha \beta \gamma \delta }\,e_\alpha ^a\,e_\beta ^b\,e_\gamma
^c\,e_\delta ^d
=\det \left| \left| e_\alpha ^a\right| \right|\, \epsilon ^{abcd}
=(\pm )\sqrt{g}\, \epsilon ^{abcd}
\quad {\rm and}\quad
\sqrt{g}\,\epsilon^{abcd}\,e_a^\alpha \,e_b^\beta \,e_c^\gamma \,e_d^\delta
=(\pm ) \epsilon^{\alpha \beta \gamma \delta }.
\label{e^alpha->e^a}
\end{equation}
Now the $(\pm )$-option is associated with the orientation of the vierbein.
We shall stick to the plus sign again.

Let us turn to properties of $\Omega \,e_\mu ^a\,e_\nu ^b\,
\hat{\xi}_{ab}^{+}\,\Omega ^{-1}$. The gauge rotation respects duality and
we may omit the $\Omega ,\,\Omega ^{-1}$ matrices. Let us calculate the
dual of $e_\mu ^a\,e_\nu ^b\,\hat{\xi}_{ab}^{+}$:
\begin{equation}
\frac 12\epsilon ^{\mu \nu \lambda \sigma }\,
e_\lambda ^a\,e_\sigma ^b\,\hat{\xi}_{ab}^{+}
=\frac 12\epsilon ^{\mu \nu \lambda \sigma }\,
\frac{\partial q^\gamma }{\partial x^\lambda }e_\gamma ^c\,
\frac{\partial q^\delta }{\partial x^\sigma }e_\delta ^d\,
\hat{\xi}_{cd}^{+}
\end{equation}
With the help of the equation (\ref{e^mu->e^alpha}) one can show that:
\begin{equation}
\frac 12\epsilon ^{\mu \nu \lambda \sigma }\,
e_\lambda ^a\,e_\sigma ^b\,\hat{\xi}_{ab}^{+}
=\frac 12\frac{\epsilon ^{\alpha \beta \gamma \delta }}{\sqrt{g}}
\frac{\partial x^\mu }{\partial q^\alpha }
\frac{\partial x^\nu }{\partial q^\beta }
e_\gamma ^c\,e_\delta ^d\,\hat{\xi}_{cd}^{+};
\end{equation}
and then using (\ref{e^alpha->e^a}) and the identity $e_a^\alpha \,e_\beta
^a=\delta _\beta ^\alpha $ one arrives at:
\begin{equation}
\frac 12\frac{\epsilon ^{\alpha \beta \gamma \delta }}{\sqrt{g}}
\frac{\partial x^\mu }{\partial q^\alpha }
\frac{\partial x^\nu }{\partial q^\beta }
e_\gamma ^c\,e_\delta ^d\,\hat{\xi}_{cd}^{+}
=\frac 12\frac{\partial x^\mu }{\partial q^\alpha }e_a^\alpha \,
\frac{\partial x^\nu }{\partial q^\beta }e_b^\beta \,
\epsilon ^{abcd}\,\hat{\xi}_{cd}^{+}
=e^{a\,\mu }\,e^{b\,\nu }\, \hat{\xi}_{ab}^{+}.
\end{equation}
This proves that $e_\mu ^a\,e_\nu ^b\,\hat{\xi}_{ab}^{+}$ is selfdual (and
$e_\mu ^a\,e_\nu ^b\,\hat{\xi}_{ab}^{-}$ is antiselfdual) provided that the
vierbein and the Cartesian coordinates are oriented in the same way.


\end{document}